\documentclass[conference]{IEEEtran}
\IEEEoverridecommandlockouts
\usepackage{amsfonts} 
\usepackage{graphicx}
\usepackage{float}
\usepackage{amsmath}
\usepackage{booktabs}
\usepackage{multirow}
\usepackage{color}
\usepackage{array}
\usepackage[ruled,norelsize,vlined,linesnumbered]{algorithm2e}
\usepackage{cite}
\usepackage{marvosym}

\makeatletter
\newcommand{\removelatexerror}{\let\@latex@error\@gobble}
\makeatother

\ifCLASSINFOpdf
\else
\fi
\begin{document}
%
\title{Financial Risk Assessment via Long-term Payment Behavior Sequence Folding}


\author{
    \IEEEauthorblockN{Yiran Qiao\IEEEauthorrefmark{1}, 
    Yateng Tang\IEEEauthorrefmark{2}, 
    Xiang Ao\IEEEauthorrefmark{1}\IEEEauthorrefmark{4},
    Qi Yuan\IEEEauthorrefmark{1},
    Ziming Liu\IEEEauthorrefmark{2},
    Chen Shen\IEEEauthorrefmark{2},
    Xuehao Zheng\textsuperscript{\Letter}\thanks{\Letter~Xuehao Zheng is the corresponding author.}\IEEEauthorrefmark{2}
    }

    \IEEEauthorblockA{\IEEEauthorrefmark{1} 
    University of Chinese Academy of Sciences, Beijing 100049, China.}
    
    \IEEEauthorblockA{\IEEEauthorrefmark{4} 
    Institute of Intelligent Computing Technology, Suzhou, CAS.}
    
    \IEEEauthorblockA{\IEEEauthorrefmark{2}
    Tencent Weixin Group, Shenzhen, China.\\
    yrqiao@gmail.com, aoxiang@ucas.ac.cn, yuanqi15@mails.ucas.ac.cn\\
    \{fredyttang, simonzmliu, achenshen, xuehaozheng\}@tencent.com
    }
}


%


\maketitle

\begin{abstract}
Online inclusive financial services encounter significant financial risks due to their expansive user base and low default costs. 
By real-world practice, we reveal that utilizing longer-term user payment behaviors can enhance models' ability to forecast financial risks.
However, learning long behavior sequences is non-trivial for deep sequential models. 
Additionally, the diverse fields of payment behaviors carry rich information, requiring thorough exploitation.
These factors collectively complicate the task of long-term user behavior modeling.
To tackle these challenges, we propose a \underline{\textbf{L}}ong-term Payment \underline{\textbf{B}}ehavior \underline{\textbf{S}}equence \underline{\textbf{F}}olding method, referred to as \textbf{LBSF}.
In LBSF, payment behavior sequences are folded based on merchants, using the merchant field as an intrinsic grouping criterion, which enables informative parallelism without reliance on external knowledge. Meanwhile, we maximize the utility of payment details through a multi-field behavior encoding mechanism.
Subsequently, behavior aggregation at the merchant level followed by relational learning across merchants facilitates comprehensive user financial representation. 
We evaluate LBSF on the financial risk assessment task using a large-scale real-world dataset. The results demonstrate that folding long behavior sequences based on internal behavioral cues effectively models long-term patterns and changes, thereby generating more accurate user financial profiles for practical applications.

\end{abstract}


%
\IEEEpeerreviewmaketitle

\section{Introduction}

In recent years, online payment platforms have promoted diverse inclusive financial services, extending their reach to a broader user base underserved by traditional financial institutions~\cite{musto2015personalized,guild2017fintech,taherdoost2023fintech}.
However, both this demographic expansion and low default costs contribute to persistently high default risk in online inclusive finance~\cite{sun2023digital,loubere2017china}. 
It burgeons the significance and urgency of the financial risk assessment task in these online payment platforms to underpin a more robust modern finance ecosystem. 

Financial risk assessment aims to evaluate the default potential of users in financial services by analyzing individuals' financial history~\cite{chen2016financial}. 
As the era evolves and users' financial histories grow, techniques that traditional financial institutions employ for user evaluation
have progressed from statistical methods~\cite{altman1968financial,dimitras1996survey,hand1997statistical} to machine learning models~\cite{baesens2003benchmarking,li2006evaluation,luo2020unsupervised}. However, what remains unchanged is the reliance on records of conventional financial activities such as credit card usage, loan repayments, and mortgage history as the foundation for data analysis~\cite{chen2016financial,crook2007recent}.
Nevertheless, online platforms usually do not possess users' private financial history records due to various restrictions.

Despite this data deprivation, online payment platforms have their unique advantages in accessing vast amounts of user payment behavior data, which can be leveraged to enhance risk management and detect high default risk. While such payment behaviors do not directly represent financial credit, these data, often in the form of sequences, can provide valuable insights into a user's financial habits, spending patterns, and overall creditworthiness. Therefore, these sequential behavioral data can be regarded as para-financial information, which can also serve as the data foundation for financial risk assessment.
%


Similar ideas could be found in some existing works of user sequential behavior modeling for financial risk assessment\cite{guo2018learning,cheng2020spatio,lin2021online,xi2020neural,lidesign2022}. 
These works follow a methodology that employs state-of-the-art sequential models as a foundation, incorporating tailored attention mechanisms~\cite{guo2018learning,cheng2020spatio,lin2021online} or hierarchy establishment through external knowledge~\cite{xi2020neural,lidesign2022} to capture contextual signals regarding users' financial well-being.
Despite their remarkable success, these attempts focus on relatively short sequences, overlooking the significance of long-term behaviors. 



\begin{figure}[H]
  \centering
  \includegraphics[width=0.8\linewidth]{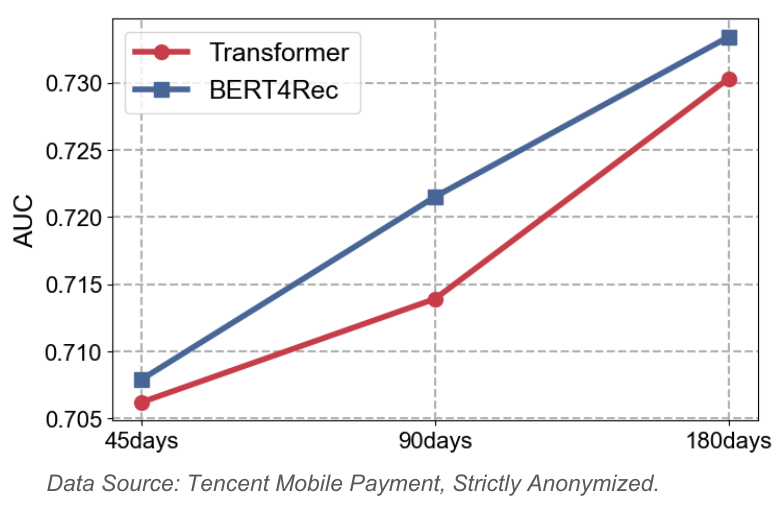}
  \caption{The performance of Transformer and BERT4Rec on real-world datasets of 45 days, 90 days, and 180 days, measured by AUC.} 
  \label{fig:intro}
\end{figure}
Meanwhile, the available user behavior sequences on online payment platforms are continually growing. In real business practice, as depicted in Figure~\ref{fig:intro},  we experimented with behavior data spanning 45, 90, and 180 days to predict defaults, and the model performance shows a monotonic increase with longer sequence lengths. 
The observed phenomenon, that long-term sequential behaviors often derive better performances of the models suggests further exploitation of abundant para-financial behaviors. 
Consequently, both the data growth and the practical insights call for the upgrade to long-term user payment behavior modeling for financial risk assessment.

However, deploying sequential models for long-term payment behaviors is challenging. The complexity escalates when dealing with extremely long sequences of user behavior data. 
Taking the widely deployed Transformer architecture as an example, time and storage costs increase quadratically with the length of the user behavior sequences~\cite{vaswani2017attention}. Given the operational problem, a straightforward solution is to divide the entire sequence into shorter segments for parallel processing. However, this segmentation may lead to the loss of valuable long-range contextual information. It is crucial to find a segmentation method that brings semantic information gain to compensate for this loss. 

Following this idea, we aim to apply an internally guided segmentation approach based on data fields within behaviors. This approach organizes data systematically to reduce chaos and allows for adaptable sequence segmentation across various scenarios. Fortunately, long-term payment behaviors originally encompass rich data fields, including textual, temporal, and numerical data, allowing segmentation based on the inherent features of the raw behaviors themselves, rather than relying on external knowledge. Furthermore, intending to maximize the data utility, we design a multi-field encoding mechanism to 
characterize user payment behaviors, thus maintaining the para-financial details to the greatest extent possible.

To this end, in this paper, we propose a method to fold the long-term user payment behavior sequences at the merchant level for financial risk assessment. We refer to the method as ``\underline{\textbf{L}}ong-term Payment \underline{\textbf{B}}ehavior \underline{\textbf{S}}equence \underline{\textbf{F}}olding'', ``\textbf{LBSF}'' in short. Specifically, we fold behavior sequences based on the merchant field of payment behaviors, grouping payment behaviors by merchants and ordering payment behaviors within the same merchant chronologically. 
Folding sequences at the merchant level not only enables segmentation for parallel processing but also provides additional information gain. 

Moreover, we encode payment behaviors by integrating all available data fields, i.e., the payment's description, timing, and amount. With such a multi-field joint encoding mechanism, payment behaviors can be represented more comprehensively,  preserving valuable details that enrich the context for long-term user behavior learning. 
For information melding within and across merchants, based on the parallelism of Transformer~\cite{vaswani2017attention}, sub-sequences at the merchant level can first be aggregated in parallel, followed by a secondary aggregation of merchant-level representations, to obtain the final user representation.
In this way, LBSF can learn long-term user behavior sequences effectively.

We conduct extensive experiments on a real-world large-scale dataset provided by Tencent Mobile Payment to validate our method's effectiveness, demonstrating LBSF's ability to comprehensively characterize user financial profiles.
Our main contributions are summarized as follows:
\begin{itemize}
    \item To our knowledge, we are the first to fold long behavior sequences based on the inherent characteristics of behaviors themselves. This reorganization of long-term behaviors empowers informative parallel processing and easy adaptation across applications.
    \item We propose the \underline{\textbf{L}}ong-term Payment \underline{\textbf{B}}ehavior \underline{\textbf{S}}equence \underline{\textbf{F}}olding~(LBSF) method for financial risk assessment.  LBSF folds long-term user payment behavior sequences at the merchant level to capture the patterns and changes of payment behaviors within and across merchants, with all available para-financial details leveraged via multi-field behavior encoding.
    \item Experiments on a large-scale real-world dataset demonstrate the effectiveness and applicability of our proposed method LBSF.
\end{itemize}

\section{Related Work}
The related research of this work can be divided into three categories: the research on financial risk assessment, sequential user behavior modeling for fraud detection, and Transformer-based methods for long-sequence modeling.

\subsection{Financial Risk Assessment}
One of the earliest attempts at financial risk assessment was made in 1968, as Altman~\cite{altman1968financial} proposed a linear model based on various financial features to predict corporate bankruptcy. It marked the beginning when traditional financial institutions set up default risk rating procedures on their own. In this stage, statistical methods based on the fundamental characteristics and financial status of users dominated the field, which was systematically reviewed by Dimitras~\cite{dimitras1996survey}. 

As the user community rapidly expanded and Artificial Intelligence technology developed, it became evident that machine learning methods could substantially enhance both the efficiency and accuracy of statistical approaches without relying on restrictive assumptions~\cite{chen2016financial}. To name some, decision tree~(DT)~\cite{baesens2003benchmarking} and support vector machine~(SVM)~\cite{li2006evaluation,luo2020unsupervised} were among the widely employed techniques, utilizing extensive user features.

However, accessing extensive user characteristics such as income, assets, and debts, which are available to traditional financial institutions, can present difficulties in online finance owing to regulatory restrictions. 
As user behavior data grows on online platforms, researchers have introduced deep learning models to model the accumulated behaviors. These models can detect different types of defaults, e.g., fraud~\cite{guo2018learning,liu2020fraud,ling2023learned,liu2021intention}, cash-out~\cite{hu2019cash,ji2022detecting}, etc~\cite{liang2021credit,zhong2020financial}. A recent comprehensive review by Zhu and Ao~\cite{zhu2021intelligent} could provide more details. 
\subsection{User Behavior Sequence Modeling for Fraud Detection}
Previous works have researched user behavior sequence modeling applied for fraud detection~\cite{guo2018learning,cheng2020spatio,lin2021online,xi2020neural,lidesign2022}.
By accounting for the order of user actions, this modeling approach enables a deeper comprehension of behavior correlations and a reflection of user intentions~\cite{ijcai2023p0746}. 

For fraud detection, Guo~\cite{guo2018learning} incorporated the self-historical attention module and interactive module for LSTM to help identify repeated or cyclical behaviors. Cheng~\cite{cheng2020spatio} proposed a spatial-temporal attention-based neural network~(STAN) for credit card fraud detection, to jointly analyze the spatial and temporal information of sequential transactions. 
While these models use tailored attention mechanisms to capture contextual signals in behavior sequences, they may overlook the hierarchical information.

Considering the hierarchical information, SAH-RNN~\cite{lin2021online} incorporated web page structures into behavior sequences and applied dual attention mechanisms to perceive users’ global and local intentions. NHFM~\cite{xi2020neural} applied event and sequence extractors as a hierarchical structure to learn event and sequence representations. SHORING~\cite{lidesign2022} consisted of the high-order interaction network and conditional sequence network to learn various symbolic expressions through sequence data. These works consider hierarchy in terms of web page structures or extracted events for additional information gain beyond the raw behaviors themselves.



Our proposed method, however, delves into the inherent nature of payment behaviors. We reorganize and learn sequential behaviors to the merchant hierarchy, inherently reflecting users' financial status and consumption habits, and accordingly fully utilize textual information in addition to transaction timings and amounts. Consequently, the merchant-based embeddings are further learned as high-level sequences for the final user representations.




\subsection{Transformer-based Long Sequence Modeling}
Since our method employs Transformers as the backbone sequential model, we summarize Transformer-based approaches for long sequence modeling as follows.
Transformers, with their innovative self-attention mechanism, have surpassed traditional sequential models like RNNs and LSTMs in efficacy and scalability~\cite{vaswani2017attention}. They have become dominant across diverse domains, including language processing~\cite{kenton2019bert,raffel2020exploring}, image analysis~\cite{han2022survey}, and protein study~\cite{rives2021biological}.
However, Transformers' quadratic time and memory complexity for input length limit their application in domains requiring longer sequences~\cite{tay2022efficient}. In response to this operating problem, researchers have made attempts to design various variants of Transformers to scale to long sequences~\cite{kitaev2019reformer,dai2019transformer,wang2020linformer,zhou2021informer}. For instance, Reformer~\cite{kitaev2019reformer}  
introduced Locality-Sensitive Hashing~(LSH) attention and reversible residual layers to reduce the memory footprint. 
Transformer-XL~\cite{dai2019transformer} proposed a segment-level recurrence mechanism that connects multiple segments.
Linformer~\cite{wang2020linformer} projected the length dimension of keys and values to a lower-dimensional representation with low-rank self-attention. 
Informer~\cite{zhou2021informer} combined ProbSparse self-attention mechanism and distilling operation to enhance the scalability of traditional Transformer architecture.

Previous studies have mainly concentrated on adapting conventional Transformers to handle general tasks regarding long sequence modeling. In contrast, our approach diverges from them by directing attention toward the intrinsic characteristics of long-term payment behavior data. Our primary goal is to develop a specialized methodology that not only tackles operational challenges but also facilitates the exploration of underlying correlations and trends within diverse, long-term payment behaviors.

\section{Problem Formulation}
In this section, we first introduce the business setting in our work and then formulate our problem in this paper.
\subsection{Business Setting}
In this work, our business setup revolves around the online inclusive financial services provided to users by an online payment platform. Our primary objective is to predict the probability that a user will commit financial default based on user long-term payment behaviors, thereby determining whether to grant them access to a variety of inclusive financial services. By accurately assessing the risk of default, the platform can make informed decisions, ensuring that only trustworthy users are granted access to certain inclusive financial services. This not only helps in safeguarding the platform's financial stability but also enhances operational efficiency and user trust.

\subsection{Problem Statement}
In this paper, we formulate our task as a binary classification problem. According to the business setting, we assign a label $y_n \in \{0, 1\}$ to each user $u_n \in \mathbf{U}$ to indicate whether they are a defaulter~$(y_n = 1)$ or not.

For each user $u_n$, a payment behavior sequence $\mathbf{s}_n=\{s_{n1},s_{n2},\cdots,s_{nT}\}$ with the length $T$ is provided, and each payment behavior $s_{ni}$ is attached with the attributes including the merchant name, the purchase description, the timing and the amount of the payment, denoted by $s_{ni} = (m_{ni},d_{ni},t_{ni},v_{ni})$. 
Given a set of labeled users on an online payment platform $\mathbf{U_\mathrm{train}}$, our goal is to predict the default labels of some unlabeled users $\mathbf{U_\mathrm{test}}=\mathbf{U} \setminus \mathbf{U_\mathrm{train}}$ with the user long-term payment behaviors provided.
Formally, the task is to learn a model $f_{\mathbf{W}}$ to classify the unlabeled users based on their long-term payment behaviors: 
\begin{equation}
        f_{\mathbf{W}}: \mathbf{\left\{s_n\right\}_{u_n \in {U}}} \rightarrow \mathbf{Y}.
\end{equation}

\section{Methodology}
\begin{figure*}[!htb]
  \centering
  \includegraphics[width=0.8\linewidth]{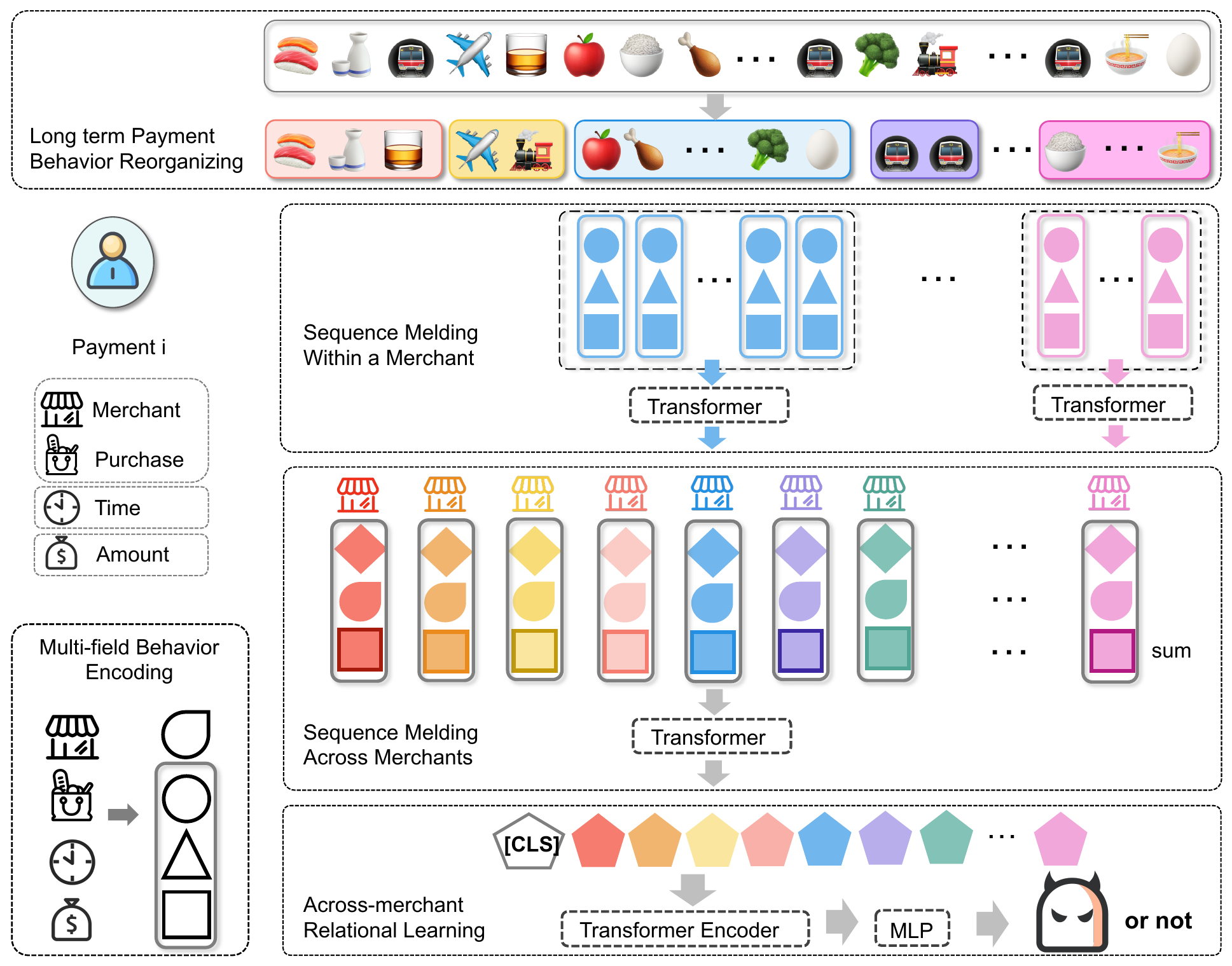}
  \caption{The architecture of LBSF.} 
  \label{fig:met:lbsf}
\end{figure*}

In this section, we detail our proposed method LBSF. First, we give an overview of LBSF. Then, we demonstrate the dissection of LBSF, including each of its integral parts:  long-term payment behavior reorganizing in Section~\ref{sec:met:fold}, multi-field behavior encoding in Section~\ref{sec:met:enc}, hierarchical sequence melding in Section~\ref{sec:met:meld}, and across-merchant relational learning in Section~\ref{sec:met:rel}.
\subsection{Overview of LBSF}
 The architecture of LBSF is illustrated in Figure~\ref{fig:met:lbsf}. 
 LBSF reorganizes the sequence of user long-term payment behaviors based on merchant categorization. Subsequently, the sequence segments for each merchant are melded hierarchically using Transformers to derive a user embedding at the single-merchant level. Finally, high-level melding and relational learning across merchants are applied to synthesize these merchant-level user embeddings into a comprehensive final user representation. This representation is further fed to a classifier for financial risk assessment.
 
\subsection{Long-term Payment Behavior Reorganizing}
\label{sec:met:fold}


To better exploit the long-range dependencies and contextual insights in long-term user payment behaviors, building a hierarchical structure for behaviors appears a reasonable way.
Divergent from previous methods that rely on external cues like web page structures~\cite{lin2021online} or events~\cite{xi2020neural,lidesign2022} for hierarchy establishment, we segment and manage the entire sequence based on the intrinsic signals present within the raw payment behaviors, i.e., the merchants. Specifically, we reorganize the whole behavior sequence according to the merchant $m_{ni}$, preserving the chronological order within each merchant. 

\begin{figure*}[htb]
  \centering
  \includegraphics[width=0.75\linewidth]{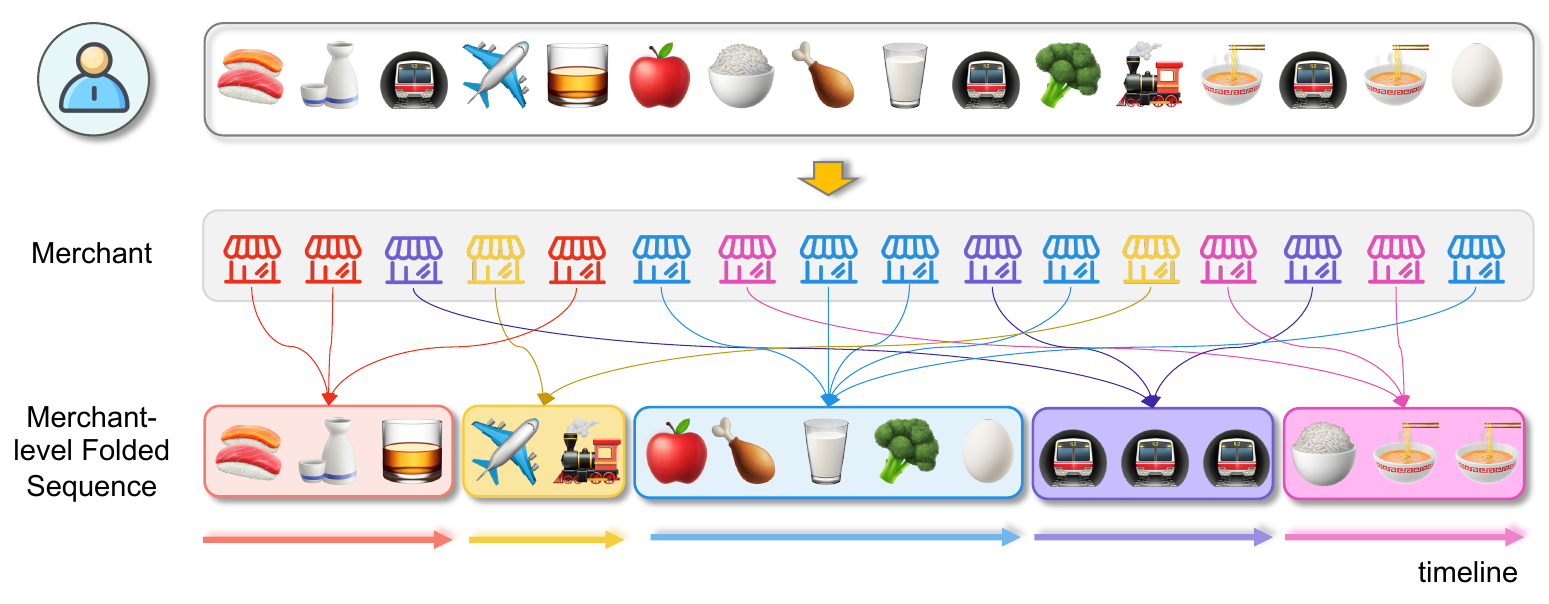}
  \caption{The illustration of long-term payment behavior reorganizing at merchant level on a short example sequence.} 
  \label{fig:met:folding}
\end{figure*}

An example process of this merchant-level reorganizing is depicted in Figure~\ref{fig:met:folding}. After the reorganization, the chaotic behavior sequence becomes well-organized, enabling some direct observations from various payment behaviors across and within merchants: The user always commuted by subway during this period. Early in the time window, he frequented an upscale restaurant and booked a flight via a travel app. But later from a certain point, he began opting for quick meals from a convenience store and buying basic groceries from a local market. Besides, through the same travel app, instead of booking flights, his new choice is cheap train tickets.

After reorganizing the behaviors, the observed changes within and across merchants can reflect the evolving spending habits. This example may imply a relatively increased likelihood of financial default, as it initially involves a period of concentrated luxury spending followed by a transition to basic consumption.
We believe such reorganizing at the merchant level can facilitate comprehending and capitalizing on trends and shifts of financial status within or across merchants.

For operation, note that the number of merchants $M$ is set. For each user $u_n$, we have $M$ merchants as $\{m^\prime_{n1},m^\prime_{n2},\cdots,m^\prime_{nM}\} $ and corresponding
$M$ merchant-level sub-sequences denoted by $\{\mathbf{s}^\prime_{n1},\mathbf{s}^\prime_{n2},\cdots,\mathbf{s}^\prime_{nM}\}$ after sequence reorganizing.
These sub-sequences are further learned by Transformers for fused embeddings at the merchant level. 

\subsection{Multi-field Behavior Encoding}
\label{sec:met:enc}
Recall that user payment behavior data encompasses diverse fields, namely merchant name, purchase description, timing, and transaction amount. The key question that arises is how to effectively encode and integrate these varied data fields.

Within a payment behavior $s_{ni} = (m_{ni},d_{ni},t_{ni},v_{ni})$, the transaction amount $v_{ni}$ is already numerical and the texts $m_{ni},d_{ni}$ can be easily encoded by Transformers.

For time embedding, in this scenario, due to the relatively sparse payment behavior sequences of users, only relying on positional embeddings based on relative positions is insufficient to capture changes in user consumption habits.  
Taking each day's 24-hour cycle as an example, the timestamps when a user paid for midnight snacks at 23:00 and 0:00 should be closely embedded in terms of a cyclic pattern. Bur plain numerical time embeddings fail to achieve this. Therefore, we intend to better represent the periodicity of each time dimension, namely month, day, week, and hour.
Considering the periodic nature of the sine/cosine function, a sine-cosine transformation-based time embedding is further proposed. Transaction time $t_{ni}$ is encoded by 
\begin{equation}
    \phi(t,T) = (\cos(2\pi t/T,\sin(2\pi t/T ),
\end{equation}
as in each dimension, e.g., in the hour dimension $T=24$. Note that month, day, week, and hour are all encoded accordingly, and the embeddings of each are concatenated to form the general time embedding.

 Here, we have transformed the heterogeneous data into the embedding space for further integration. Note that the embeddings of $d_{ni},t_{ni}$ are denoted by $\mathbf{d}_{ni},\mathbf{t}_{ni}$ for simplicity. The embedding of a single behavior $\mathbf{e}_{ni}$ is obtained by concatenation of $\mathbf{d}_{ni},\mathbf{t}_{ni},v_{ni}$ and dimensionality reduction through a fully connected layer.
 For user $u_n$, $\{\mathbf{m}^\prime_{n1},\mathbf{m}^\prime_{n2},\cdots,\mathbf{m}^\prime_{nM}\}$ represents the text embeddings of $M$ merchants  $\{m^\prime_{n1},m^\prime_{n2},\cdots,m^\prime_{nM}\} $.

\subsection{Hierarchical Sequence Melding}
\label{sec:met:meld}
Next, we elaborate on how to meld sub-sequences within and across merchants through Transformers respectively.
\subsubsection{Melding within a merchant}
For a user $u_n$ and a merchant $m^\prime_{nj} \in \{m^\prime_{n1},m^\prime_{n2},\cdots,m^\prime_{nM}\} $ patronized by $u_n$, the merchant-level sub-sequence of behavior embeddings is denoted as $\mathbf{s}^\prime_{nj} = \{\mathbf{e}_{nk}\}_{m_{nk} = m^\prime_{nj}}  $.
These embeddings are fed into a Transformer and then processed via average pooling to obtain a representation within the merchant $\mathbf{h}^M_{nj}$ as in:
\begin{equation}
    \mathbf{h}^M_{nj} = \text{AveragePooling}(\text{Transformer}(\mathbf{s}^\prime_{nj}))
    \label{eq:mwm}
\end{equation}

In this way, the periodic patterns and consumption habits within the same merchant's payment behaviors can be better extracted. Then, the merchant representation of payment behaviors $\mathbf{h}^M_{nj}$ is further fused with the text embedding of the merchant $\mathbf{m}^\prime_{nj}$ through a fully connected layer, producing the merchant embedding still denoted by $\mathbf{h}^M_{nj}$.
With the fusion of semantic information, the traits and the tiers of the merchants are integrated into the merchant embeddings.

\subsubsection{Melding across merchants}
After melding payment behaviors within each merchant in a parallel manner, for user $u_n$, we have a sequence of merchant embeddings: $\mathbf{h}^M_n = \{\mathbf{h}^M_{n1},\mathbf{h}^M_{n2},\cdots,\mathbf{h}^M_{nM}\}$.
Similar to melding within merchants, we utilize Transformer to learn across merchants to generate enhanced merchant embeddings :
\begin{equation}
    \mathbf{h}^{M \prime}_{n} = \text{Transformer}(\mathbf{h}^{M}_{n})
       \label{eq:mam}
\end{equation}
Consequently, we obtain a sequence of merchant embeddings $\mathbf{h}^{M\prime}_n = \{\mathbf{h}^{M\prime}_{n1},\mathbf{h}^{M\prime}_{n2},\cdots,\mathbf{h}^{M\prime}_{nM}\}$ with more contexts and dependencies fused through the self-attention of Transformer.
\subsection{Across-merchant Relational Learning}
\label{sec:met:rel}
Across-merchant relational learning refers to understanding the correlations among different merchant embeddings. This process assigns weights to different merchants, leading to comprehensive user representations that merge merchant representations more effectively.

Inspired by Vision Transformer~(ViT)~\cite{dosovitskiy2020image}, we consider the merchant embeddings extracted from a user payment behavior sequence as the patch embeddings partitioned from an image.
A ViT-like self-attention mechanism is applied to accomplish across-merchant relational learning. 

One key distinction of ViT compared to vanilla Transformers is the inclusion of a classification token, represented by $\text{[CLS]}$. This token $\text{[CLS]}$ is integrated into the Transformer Encoder alongside the embeddings of individual merchants $ \{\mathbf{h}^{M\prime}_{n1},\mathbf{h}^{M\prime}_{n2},\cdots,\mathbf{h}^{M\prime}_{nM}\}$. During the self-attention process, the classification token attends to the merchant embeddings, allowing the model to capture global contextual information across all merchants in the sequence.

After the self-attention process, the classification token is fused with the other merchant embeddings. This fusion process enables the model to aggregate information both locally, by considering individual merchants, and globally, by incorporating the global context represented by the classification token. 

Ultimately, the embedding $\mathbf{h}^\prime_n$ at $\text{[CLS]}$ serves as a comprehensive representation of the entire input merchant embedding sequence, i.e., the final representation of user $u_n$. Utilizing a Multi-Layer Perceptron (MLP) as the classifier, the prediction of the user credit default is then accomplished.

\begin{equation}
    \hat{y}_n = \text{MLP}(\mathbf{h}^\prime_n)
    \label{eq:crl}
\end{equation}

We perform model training by minimizing the  Binary Cross-Entropy Loss~(BCELoss), with $N$ as the number of training samples:
\begin{equation}
    \mathcal{L}(y, \hat{y}) = -\frac{1}{N} \sum_{n}^{N} [y_n \log(\hat{y}_n) + (1 - y_n) \log(1 - \hat{y}_n)]
    \label{eq:loss}
\end{equation}

In summary, the training process of LBSF is illustrated in Algorithm~\ref{alg:lbsf}.

\begin{algorithm}
\caption{LBSF Training}
\label{alg:lbsf}
\KwIn{Training dataset $\{(\mathbf{s}_n, y_n)\}_{n=1}^N$, where $\mathbf{s}_n=\{s_{n1},s_{n2},\cdots,s_{nT}\}$ is the behavior sequence with each behavior $s_{ni} = (m_{ni},d_{ni},t_{ni},v_{ni})$.}
\KwOut{Trained model parameters $\mathbf{W}$. }

  \For{$n = 1, \ldots,N$}{
  Reorganize $\mathbf{s}_n$ at the merchant level for sub-sequences~$\{\mathbf{s}^\prime_{n1},\mathbf{s}^\prime_{n2},\cdots,\mathbf{s}^\prime_{nM}\}$ of $M$ merchants selected $\{m^\prime_{n1},m^\prime_{n2},\cdots,m^\prime_{nM}\}$.
  
    \For{merchant~$m^\prime_{nj}  \in \{m^\prime_{n1},m^\prime_{n2},\cdots,m^\prime_{nM}\}$}{
    Encode merchant text $m^\prime_{nj}$ as $\mathbf{m}^\prime_{nj}$.
    
Encode each behavior ${s}_{nk} \in \mathbf{s}^\prime_{nj}$ as $\mathbf{e}_{nk}$. 

Meld behaviors within the merchant $\mathbf{s}^\prime_{nj} = \{\mathbf{e}_{nk}\}_{m_{nk} = m^\prime_{nj}} $ for $\mathbf{h}^M_{nj}$ w.r.t. Eq.~\ref{eq:mwm}.

Update $\mathbf{h}^M_{nj}$ with concatenating $\mathbf{m}^\prime_{nj}$.
    }
    Meld embeddings across merchants $\mathbf{h}^M_n = \{\mathbf{h}^M_{n1},\mathbf{h}^M_{n2},\cdots,\mathbf{h}^M_{nM}\}$ for $\mathbf{h}^{M\prime}_n = \{\mathbf{h}^{M\prime}_{n1},\mathbf{h}^{M\prime}_{n2},\cdots,\mathbf{h}^{M\prime}_{nM}\}$ w.r.t. Eq.~\ref{eq:mam}.
    
    Perform across-merchant relational learning for user representation $\mathbf{h}^\prime_n$ w.r.t. Eq.~\ref{eq:crl}.

  }

    Train the model by minimizing Eq.~\ref{eq:loss}.
   
     Return $\mathbf{W}$.
\end{algorithm}

\section{Experimental Setup}
In this section, we present the details of our experimental setup, regarding the datasets, compared methods, implementation details, and metrics.
\subsection{Datasets}
We obtain a real-world financial risk assessment dataset from Tencent Mobile Payment\footnote{The dataset in this paper is properly sampled only for testing purposes and does not imply any commercial information. All users' private information is removed from the dataset. Besides, the experiment was conducted locally on Tencent's internal server by formal employees who strictly followed data protection regulations.}. This large-scale dataset collected the long-term payment behaviors of 458,744 users.

For data splits, the dataset includes 387,332 users for training (from March 24, 2020, to May 19, 2021), 36,021 users for validation (from May 20, 2021, to May 31, 2021), and 35,391 users for testing (from June 1, 2021, to June 16, 2021). For label annotation, we define users who default after being granted access to inclusive financial services as positive samples; otherwise as negative samples. Note that this dataset is sampled from the user corpus to maintain a positive rate among all included users of approximately 10.0\%. The statistics of the dataset are presented in Table \ref{tab:exp:dataset}.

\begin{table}[b]
\caption{The statistics of Tencent Mobile Payment dataset.}
\centering
\begin{tabular}{c|rrrc}
\toprule
Dataset & \#Pos. & \#Neg. & \#Total & \%Pos. Rate \\
\midrule
Training & 27,836 & 359,496 & 387,332 & 7.2\% \\

Validation & 4,378 & 31,643 & 36,021 & 12.2\% \\

Testing & 4,330 & 31,061 & 35,391 & 12.2\% \\
\bottomrule
\end{tabular}
\label{tab:exp:dataset}
\end{table}

\subsection{Compared Methods}
To adequately demonstrate the effectiveness of our proposed method LBSF, we compared LBSF with 13 baselines, categorized into three types: i.e., basic models, sequential models, and pre-training models.
\subsubsection{Basic Models}

XGBoost~\cite{chen2016xgboost} and MLP~\cite{rosenblatt1958perceptron} are two widely used machine learning algorithms that we experimented with based on user statistical features.

\subsubsection{Sequential Models}

BiLSTM~\cite{zhang2015bidirectional} is a long short-term memory network that processes input sequences in both forward and backward directions. Transformer~\cite{vaswani2017attention} is one of the most widely used sequential models. Various Transformer variants are proposed for scalability, among which we compare: Longformer~\cite{beltagy2020longformer} tailors Transformer with global attention mechanisms and sliding windows. Reformer~\cite{kitaev2019reformer} is a Transformer variant with locality-sensitive hashing attention. Informer~\cite{zhou2021informer} adapts Transformer to scale to long time series with Prob-Sparse attention and distillation. In addition to Transformer variants, 
STAN~\cite{cheng2020spatio} is selected as an attention-based sequential model for credit card fraud detection.

Besides,
we also experimented with several newly proposed Transformer equivalents:  RetNet~\cite{sun2023retentive} introduces a retention mechanism for improving sequence modeling. RWKV~\cite{peng2023rwkv} combines the efficient parallelizable training of Transformers with the efficient inference of RNNs. Mamba~\cite{gu2023mamba} integrates selective structured state space models into a simplified end-to-end neural network architecture without attention, scalable to million-length sequences.


\begin{table*}[h]
\caption{Performance Comparison of LBSF with baselines of three types, namely basic models, sequential models, and pre-training models. The best results are bolded and the second best are underlined.}
\centering
\begin{tabular}{c|c|lc|lc|lc}
\toprule
\multirow{2}{*}{Type} &\multirow{2}{*}{Model}
& \multicolumn{2}{c|}{45~days} &\multicolumn{2}{c|}{90~days} &\multicolumn{2}{c}{180~days} \\
& & AUC & Recall@10\% & AUC & Recall@10\% & AUC & Recall@10\% \\
\midrule
\multirow{2}{*}{Basic} 
& XGBoost & \multicolumn{2}{c|}{——} &0.6774&0.2556 &\multicolumn{2}{c}{——}  \\
& MLP &  \multicolumn{2}{c|}{——} &0.6841&0.2582 &\multicolumn{2}{c}{——}  \\
\midrule
\multirow{7}{*}{\shortstack{Sequential}} 
& BiLSTM & 0.6901 &0.2632  &0.6914&0.2711&0.7011&0.2794  \\
& Transformer & 0.7062 & 0.2697  &0.7139&0.2808&0.7303&0.2931 \\
& Longformer & 0.7203 &0.2878  &0.7357&0.3023&0.7360&0.3002  \\
& Reformer & 0.7036 &0.2596  &0.7091&0.2760&0.7120& 0.2672 \\
& Informer & 0.7109 &0.2813  &0.7183&0.2769&0.7230&0.2783  \\
& STAN & 0.7105 & 0.2695 &0.7201&0.2783&0.7177&0.2760 \\
& RetNet & 0.7216 &0.2861  &0.7373&0.3016&0.7513&0.3141  \\
& RWKV & 0.7171 &0.2843  &0.7241&0.2811&0.7448&0.3023  \\
& Mamba & \textbf{0.7286} & \textbf{0.2926} &\underline{0.7449}&\underline{0.3109}&\underline{0.7554}&\underline{0.3208}  \\

\midrule
\multirow{2}{*}{Pre-training} 
& BERT4Rec &0.7079  &0.2672 &0.7215&0.2838&0.7334&0.3046   \\
& S3-Rec &0.7131  &0.2808   &0.7280&0.2917&0.7333&0.3005 \\
\midrule
Ours&\textbf{LBSF}&\underline{0.7240} &\underline{0.2924} &\textbf{0.7575}&\textbf{0.3259}&\textbf{0.7678}&\textbf{0.3467} \\
\bottomrule
\end{tabular}
\label{tab:exp:main}
\end{table*}
\subsubsection{Pre-training Models}

BERT4Rec~\cite{sun2019bert4rec} is a well-known pre-training model for user sequential behavior modeling. S3-Rec~\cite{zhou2020s3} is a pre-training method for sequential modeling that utilizes the intrinsic data correlation as self-supervision signals. These two methods are also compared due to their strong capability to model user sequential behaviors.

\subsection{Implementation Details}
All experiments are conducted on a server with A100 GPU. 
Our LBSF model is implemented using Pytorch~\cite{paszke2019pytorch}. For the optimizer, we apply the AdamW Optimizer~\cite{loshchilov2018fixing} with the learning rate 2e-4. For parameters, we adopt Transformers of one layer, with the dimension of features and the number of heads as 128 and 4 respectively. 
Besides, we set the batch size to 256, and the number of epochs to 10.
Regarding the specific hyperparameter $M$, representing the number of merchants pre-selected for each user, we set $M$ to 74.

\subsection{Metrics}
We adopt \textbf{AUC} and \textbf{Recall@10\%} as the evaluation metrics. AUC refers to the Area Under the Receiver Operating Characteristic~(ROC) Curve, providing an overall measure of the model's discriminative ability across all thresholds. 
Recall@10\%  is the proportion of actual positive instances correctly identified within the top 10\% ranked predictions. 

In the financial risk assessment task, the positive samples, i.e., the users who committed default, are typically a minority. Therefore, we not only use AUC to evaluate the overall discriminative ability of the models but also adopt Recall@10\% to measure the quality of the highest-ranked predictions.

\section{Experimental Results}
In this section, we analyze our experimental results on real-world datasets to demonstrate the efficacy of LBSF. Particularly, we aim to address the following research questions:
\begin{itemize}
    \item \textbf{RQ1}: Does LBSF outperform various baselines?
    \item \textbf{RQ2}: How does each component of LBSF contribute to the performance enhancement?
    \item \textbf{RQ3}: What does the merchant-level information reveal about the default users specifically identified by LBSF?
\end{itemize}
\subsection{Performance Comparison~(RQ1)}

We compare LBSF with various baselines on three subsets of varying periods (45 days, 90 days, and 180 days) from the Tencent Mobile Payment dataset to answer the first research question. These subsets consist of the same users but capture behaviors over time windows of different lengths.
The AUC and Recall@10\% scores are reported in Table~\ref{tab:exp:main}. Note that these two evaluation metrics exhibit a consistent trend. Our observations are as follows.

\subsubsection{Comparison with Baselines}
Firstly, our proposed model LBSF consistently outperforms all baseline models on all subsets of varying time periods with only one exception: On the 45-day subset, Mamba achieves only slightly higher scores than LBSF by a margin of 0.46\%, and our method still ranks second best among all models. Despite this exception, the effectiveness of LBSF is proven on real-world datasets of different lengths, indicating its applicability in business practice to identify potential defaulters.

Moreover, across the types, the sequential models have shown generally more competitive performance than the basic and pre-training methods. 
For the basic and pre-training methods, this may be attributed to the subpar user profiles generated by statistical features or general proxy tasks, which fail to capture the specific patterns in user spending activities.

Among the sequential models, recently proposed Transformer equivalents including RetNet, RWKV, and Mamba have demonstrated significant promise in long sequence modeling, outperforming variants (the X-formers) that heavily depend on the basic architecture of vanilla Transformers. Given their leading performance, these up-to-date sequential models could also act as the backbone of our LBSF to enhance its potential even further, since our folding method is applicable to all end-to-end sequential models.
\subsubsection{Comparison of Sequence Length}
Secondly, regarding the dimension of period length, models trained with longer-period payment behaviors achieve better performances. Across all the models, there is a clear trend of increasing performance with longer-period payment behavior sequences. The widely evident significance of long-term data underscores our initial premise that leveraging long-term user payment behaviors is crucial for effectively assessing financial risks, thereby enhancing the robustness and stability of inclusive finance.

The main experimental results have demonstrated the effectiveness and potential of LBSF, leading us to further investigate the contribution of each component of LBSF.

\subsection{Ablation Study~(RQ2)}
To answer the second research question, we conducted an ablation study to show that every component of LBSF contributes to the performance increase on the 90-day subset. For ablation, we subdivide the specific designs of LBSF into four components, namely merchant folding, payment description embedding, payment timing embedding, and payment amount embedding, which are removed respectively in experiments. 

Table~\ref{tab:exp:ab} displays the results of the ablation study of LBSF on the Tencent Mobile Payment dataset. The complete LBSF achieves the highest performance, affirming the effectiveness and necessity of our specialized designs for exploiting para-financial information, i.e., long-term user payment behaviors.
To remove merchant folding, we eliminate the merchant-based hierarchical structure and do not use merchant text embeddings, but keep the payments still chunked by merchants and then concatenated as a whole for input. We find out that the ablation of merchant folding leads to the largest performance degradation. This empirical finding aligns with our motivation, that the payment information at the merchant level can considerably reveal the patterns and changes in users' financial status and consumption habits, and should be explicitly highlighted in financial risk assessment.

From payment embeddings, removals of different aspects are similar: As claimed in Section~\ref{sec:met:enc}, a single behavior embedding is first obtained by concatenating the description embedding, the time embedding, and the amount embedding. Ablating the description or time or amount means excluding it in the concatenation. 
We can observe from the results that among the three aspects of payment behaviors, excluding the amount in the initial payment embedding results in the highest decrease in performance, which prompts us to prioritize payment amounts in future attempts in real-world practices.

\begin{table}[]
\caption{Ablation Study of LBSF on the 90-day Subset of Tencent Mobile Payment Dataset.}
\centering
\begin{tabular}{c|cc}
\toprule
Model & AUC & Recall@10\%  \\
\midrule

LBSF w/o merchant folding&0.7063&0.2970\\
LBSF w/o payment amount&0.7294&0.3021\\
LBSF w/o payment timing&0.7310&0.3109\\
LBSF w/o payment description &0.7375&0.3181\\
LBSF &0.7575&0.3259\\
\bottomrule
\end{tabular}
\label{tab:exp:ab}
\end{table}

\subsection{Case Study~(RQ3)}

To answer the third research question, we present and analyze two representative defaulters spotted by LBSF. For each detected defaulter, we dig into their spending habits through the attention scores assigned to merchants, as produced and utilized in the across-merchant relative learning module detailed in Section~\ref{sec:met:rel}. 

Recall that the attention scores of merchants represent the weights at which merchant embeddings contribute to the final user embedding, indicating the extent to which each merchant influences the final classification.
By extracting the weights of merchants, we demonstrate that merchant-level folding can not only help detect defaulters overlooked by baseline methods but also provide pattern-like explanations for real-world operations. Using anonymized examples, referred to as Alice and Bob, we illustrate how specific spending habits observed across various merchants may be linked to defaults.

Specifically, we selected several top-ranked merchants based on their weights and counted the number of payments at these merchants over a certain period. By observing the trends in transaction frequency across different merchants, we aim to uncover the underlying long-term patterns of spending habits and financial well-being.

\subsubsection{Defaulter Alice}
\begin{figure}[tb]
  \centering
  \includegraphics[width=0.85\linewidth]{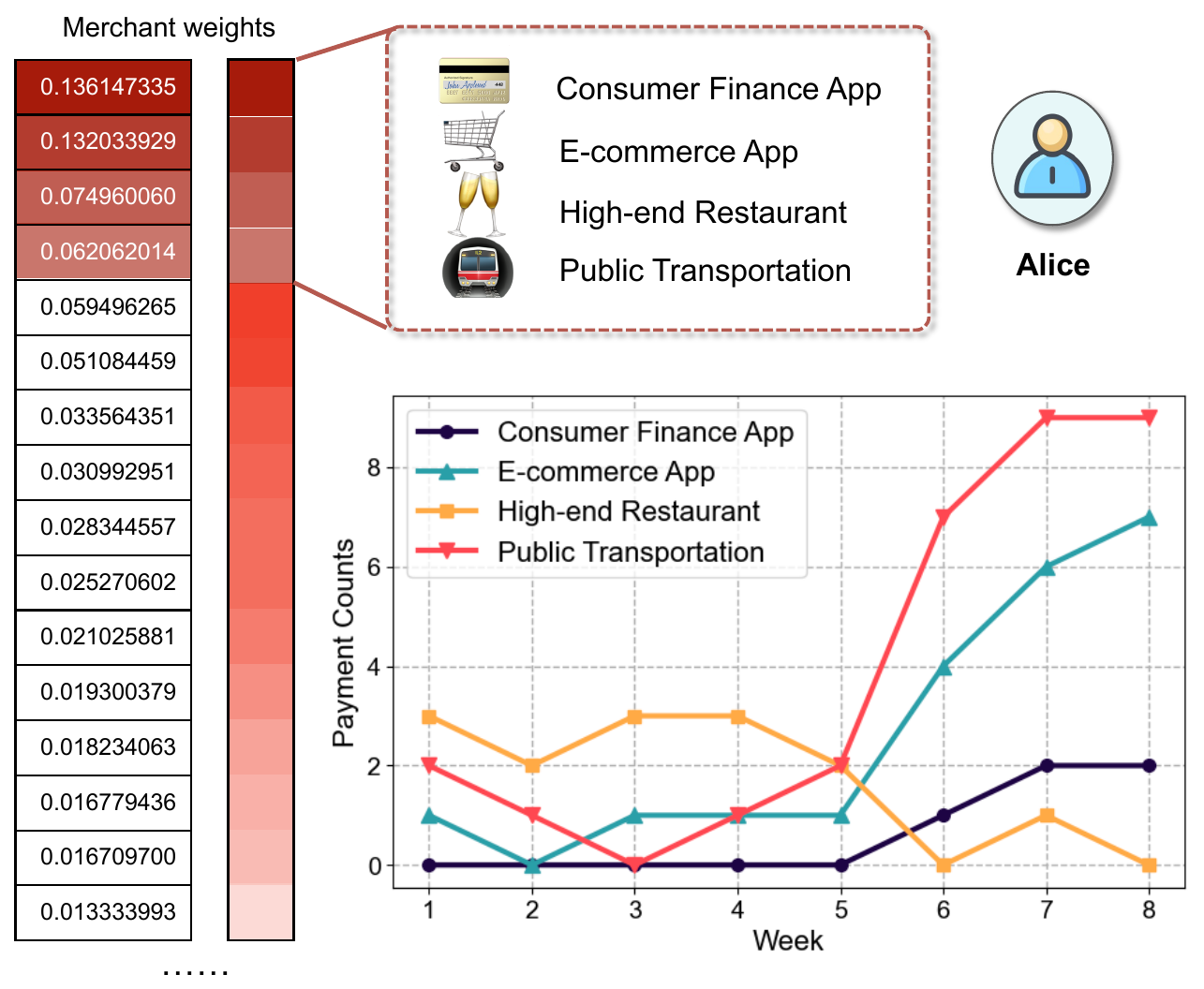}
  \caption{Case study of defaulter Alice, whose four top-ranked merchants are a consumer finance app, an e-commerce app, a high-end restaurant, and public transportation.} 
  \label{fig:res:casea}
\end{figure}
For Alice, Figure~\ref{fig:res:casea} illustrates the merchant weights and changes in consumption frequency over time at the four merchants with the highest weights.
In the line graph, the horizontal axis represents time in weekly intervals, while the vertical axis indicates the number of payments. Each marker represents the number of transactions within that week.

The four top-ranked merchants include a consumer finance app, an e-commerce app, a high-end restaurant, and public transportation. Note that in practice, we preprocess some similar merchants into one category; for example, several upscale restaurants are grouped as one.

Over the eight-week period we showcase, Alice's spending at high-end restaurants decreased while her frequency of online shopping and taking public transportation for daily needs notably increased.
Such a shift in payment behaviors became particularly evident in the fifth week when she began using consumer finance services. This adoption coincided with a sharp increase in online shopping, which may be financially supported by the services, likely attributable to financial pressures stemming from her previous high-end expenses.

Her changes in spending patterns across merchants suggest poor financial management and savings discipline. Similar trends in payment behavior should be a focal point in financial risk assessment, indicating ongoing financial strain and potential difficulty in meeting future obligations.

Note that detecting this trend may be challenging without merchant-level folding, given the original sequence's disorderliness, while LBSF can extract users' intentions across merchants with internal timelines preserved, empowering identifying potential defaulters with the similar trends.

\subsubsection{Defaulter Bob}
\begin{figure}[t]
  \centering
  \includegraphics[width=0.85\linewidth]{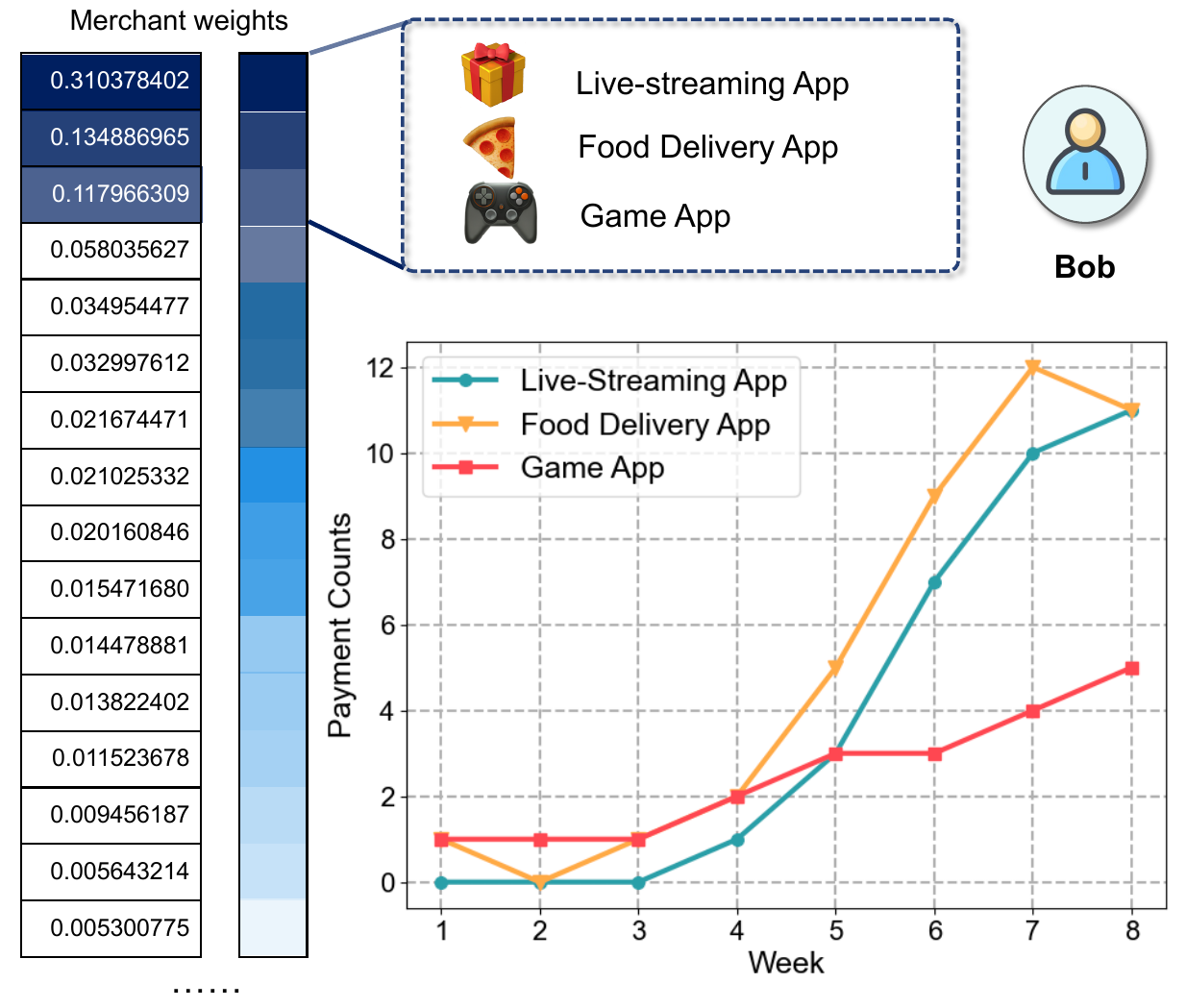}
  \caption{Case study of defaulter Bob, whose three top-ranked merchants are a live-streaming app, a food delivery app, and a game app.} 
  \label{fig:res:caseb}
\end{figure}
Similarly, Figure~\ref{fig:res:caseb} displays another case, defaulter Bob identified by LBSF.
For Bob, we select the top three merchants to analyze his default.

Bob enjoys online activities, as his top three merchant representatives include a live-streaming app, a food delivery app, and a game app. In the eight-week period we tracked, Bob's use of the aforementioned apps for live-streaming tips, food orders, and game purchases continually increased. Particularly noteworthy was the surge in his spending on tipping some online streamers, which started from 0 in the third week and peaked at shockingly 11 transactions in one week.

These payment activities of Bob have rapidly increased in frequency, indicating a trend toward impulsive spending. This tendency of emotional consumption suggests a shift toward a more extravagant lifestyle, which could jeopardize his financial stability and lead to a higher risk of financial default. This observation may prompt us to focus more on frequent emotional spending in user long-term payment behaviors which may imply an addictive tendency.

The study of two interesting cases in experiments not only demonstrates the effectiveness and interpretability of LBSF but also inspires insights into user payment behavior patterns that can be leveraged in real-world financial risk assessment. 

\section{Conclusion}
In this paper, we proposed the Long-term Payment Behavior Sequence Folding~(LBSF) method.
LBSF folds payment behavior sequences using the merchant as an intrinsic criterion and employs multi-field behavior encoding for diverse payment details.
By aggregating payment behavior sub-sequences at the merchant level and performing a secondary aggregation of merchant-level representations, LBSF effectively captures long-term patterns and changes in user financial behaviors, facilitating the characterization of user financial profiles. Our evaluation on a large-scale real-world dataset demonstrates LBSF's efficacy in leveraging long-term data to enhance financial risk assessment.




\section*{Acknowledgment}
This work is sponsored by the Tencent Rhino-Bird Focused Research Program. The research work is also supported by National Key R\&D Plan No. 2022YFC3303303, the Project of Youth Innovation Promotion Association CAS, Beijing Nova Program 20230484430, and the Innovation Funding of ICT, CAS under Grant No. E461060. 




%

\bibliographystyle{IEEEtran}
\bibliography{references}{}

\end{document}